\documentclass[%
 reprint,
 amsmath,amssymb,
 aps,
pra,
]{revtex4-2}

\usepackage{graphicx}% Include figure files
\usepackage{dcolumn}% Align table columns on decimal point
\usepackage{bm}% bold math
\usepackage{braket}

\begin{document}

\preprint{APS/123-QED}

%\title{Towards Ultrafast Quantum State Tomography using \\ Nanophotonic Parametric Amplifiers}% Force line breaks with \\
\title{Ultrafast All-Optical Measurement of Squeezed Vacuum in a Lithium Niobate Nanophotonic Circuit}
%\thanks{A footnote to the article title}%

\author{James Williams$^{1,\dagger}$, Elina Sendonaris$^{2,\dagger}$, Rajveer Nehra$^{1,3,4,5}$,\\ Robert M Gray$^1$, Ryoto Sekine$^1$, Luis Ledezma$^1$, Alireza Marandi$^{1,2}$}
\email{marandi@caltech.edu}
\affiliation{
\\ ~ \\
$^1$ Department of Electrical Engineering\text{,} California Institute of Technology\text{,} Pasadena\text{,} California 91125 \\ $^2$Department of Applied Physics\text{,} California Institute of Technology\text{,} Pasadena, California 91125 \\ $^3$Department of Electrical and Computer Engineering, University of Massachusetts Amherst\text{,} Amherst\text{,} Massachusetts 01003\text{,} USA \\ $^4$ Department of Physics\text{,} University of Massachusetts Amherst\text{,} Amherst\text{,} Massachusetts 01003\text{,} USA \\ $^5$ College of Information and Computer Science\text{,} University of Massachusetts Amherst\text{,} Amherst\text{,} Massachusetts 01003\text{,} USA\\
$^\dagger$These authors contributed equally.
}

\begin{abstract}

Squeezed vacuum, a fundamental resource for continuous-variable quantum information processing, has been used to demonstrate quantum advantages in sensing, communication, and computation. While most experiments use homodyne detection to characterize squeezing and are therefore limited to electronic bandwidths, recent experiments have shown optical parametric amplification (OPA) to be a viable measurement strategy. Here, we realize OPA-based quantum state tomography in integrated photonics and demonstrate the generation and all-optical Wigner tomography of squeezed vacuum in a nanophotonic circuit. We employ dispersion-engineering to enable the distortion-free propagation of femtosecond pulses and achieve ultrabroad operation bandwidths, effectively lifting the speed restrictions imposed by traditional electronics on quantum measurements with a theoretical maximum clock speed of 6.5 THz. We implement our circuit on thin-film lithium niobate, a platform compatible with a wide variety of active and passive photonic components. Our results chart a course for realizing all-optical ultrafast quantum information processing in an integrated room-temperature platform.
\end{abstract}

%\keywords{Suggested keywords}%Use showkeys class option if keyword
                              %display desired
\maketitle

%\tableofcontents

\begin{figure*}[th!]
\includegraphics[width=\textwidth]{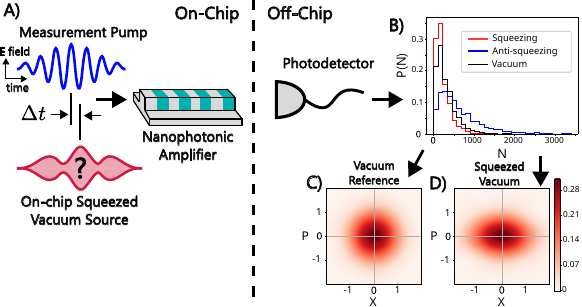}
\caption{\label{fig:fig_1} A) Layout of the measurement procedure. $\Delta t$ represents the relative time-delay between the pump and squeezed vacuum generated on-chip where a 0.775 fs delay corresponds to a measurement phase of $\phi = \frac{\pi}{2}$. B) Measured photon number distributions for squeezing ($\phi = \frac{\pi}{2}$), anti-squeezing ($\phi = 0$) and vacuum. C) Wigner function recovered for vacuum. D) Wigner function recovered for squeezed vacuum.}
\end{figure*}

\section{\label{sec:intro} Introduction}

Many quantum systems have been used to gain an advantage over otherwise purely classical means in a variety of fields \cite{degen2017quantum, gisin2007quantum, horowitz2019quantum}. Photonics has emerged as a front-runner platform in quantum information processing (QIP) for several key reasons. Most photonic technologies are capable of operating at room temperature outside a carefully-controlled cryogenic environment. Advances in integrated photonics have allowed many devices and circuits to be combined into a single monolithic platform similar to CMOS technology and the advent of integrated circuits \cite{thylen2014integrated}. Photonics also offers an inherently broad bandwidth which, when combined with dispersion engineering, can allow for the manipulation and propagation of ultra-short pulses of light \cite{ledezma2022intense}. Time-multiplexing, a pulse-based encoding technique used in photonic systems, can leverage femtosecond pulses to scale information density and throughput, and can exceed clock speeds beyond what is currently possible with conventional electronics \cite{leefmans2022topological, li2025all, bai2023photonic}.

A crucial part of any QIP system is a measurement device capable of characterizing a component of interest. Homodyne detection is widely employed as its phase-sensitive nature can be used to reconstruct the Wigner function, a quasi-probability distribution over two non-commuting variables which completely characterizes a quantum state \cite{asavanant2017generation}. Homodyne detectors can also isolate a single mode of a state of interest by shaping the spectral profile of the local oscillator appropriately \cite{roslund2014wavelength}. This phase-sensitive and mode-selective behavior makes them well-suited for characterizing a variety of quantum states, including non-classical states. Single-photon and photon number-resolving (PNR) detectors are also common tools for state characterization and have been used to generate and measure states with Wigner negativity \cite{nehra2019state, laiho2010probing, takase2021generation, gerrits2010generation, yoshikawa2017purification, sychev2017enlargement}. While homodyne and PNR detection are powerful techniques for probing quantum states of interest, their speed is ultimately constrained by the bandwidth of the electronics used to physically realize these detectors, which are limited to the GHz range.

To overcome the speed limitations of electronics, we demonstrate chip-scale all-optical Wigner tomography of squeezed vacuum (SV) using a nanophotonic optical parametric amplifier (OPA). In our previous work \cite{nehra2022few}, we measured the average photon number at the output of an OPA to calculate squeezing levels. In this work, we use a fast photodetector to measure each pulse and resolve the statistical information necessary to recover the Wigner function of the input SV. We design our OPA to have low dispersion at both the pump and signal wavelengths, allowing for the distortion-free propagation and amplification of femtosecond pulses. Such an OPA-based circuit supports a maximum measurement repetition rate (i.e. clock speed) of 6.5 THz. Increased clock speeds offers crucial benefits for QIP in photonic time-multiplexed systems as faster clocks lead to shorter processing times while also increasing the maximum circuit size that can be realized for a given system \cite{asavanant2019generation, larsen2019deterministic, madsen2022quantum}. Our results highlight nanophotonic OPAs as an important building block for chip-scale ultrafast quantum information processing systems at room temperature.

\section{\label{sec:opasas} Parametric Amplifiers as Quantum Measurement Devices}

%Talk about the history starting from the 80s paper, furusawa, Maria, other examples
Early proposals of OPAs as quantum measurement devices demonstrated loss-tolerance and detector inefficiency mitigation \cite{caves1981quantum}. OPAs are particularly well-suited for this task as their phase-sensitive amplification is in principle noiseless unlike phase-insensitive amplifiers such as erbium-doped fibers and semiconductor gain media \cite{marino_effect_2012}. OPA-assisted balanced homodyne detection has been used to characterize fiber-based sources of SV over 43 GHz of electronic bandwidth \cite{inoue2023}. A similar technique has also been demonstrated using the $\chi^{(3)}$ nonlinearity \cite{shaked2018lifting}. Other experiments have shown quantum-enhanced sensing using an SU(1,1) interferometer constructed from two OPAs for state readout \cite{Frascella_19, hudelist_quantum_2014}. OPAs have also been used to amplify weak spatially-varying signals and detect quantum correlations for imaging applications \cite{marable_measurement_1998, sun2019detection}. All-optical feed-forward, a technique used to surpass the constraints of electronics in feed-forward schemes for quantum information processing, has been demonstrated using three OPAs \cite{yamashima2024all}. Wigner-tomography of squeezed states using bulk OPAs has been experimentally realized \cite{kalash_wigner_2023}. OPA-based techniques have also been extended to squeezing across multiple spatial modes where direct detection is used to disentangle and analyze each mode independently \cite{barakat2025simultaneous}.

\begin{figure*}[t]
\includegraphics[width = \textwidth]{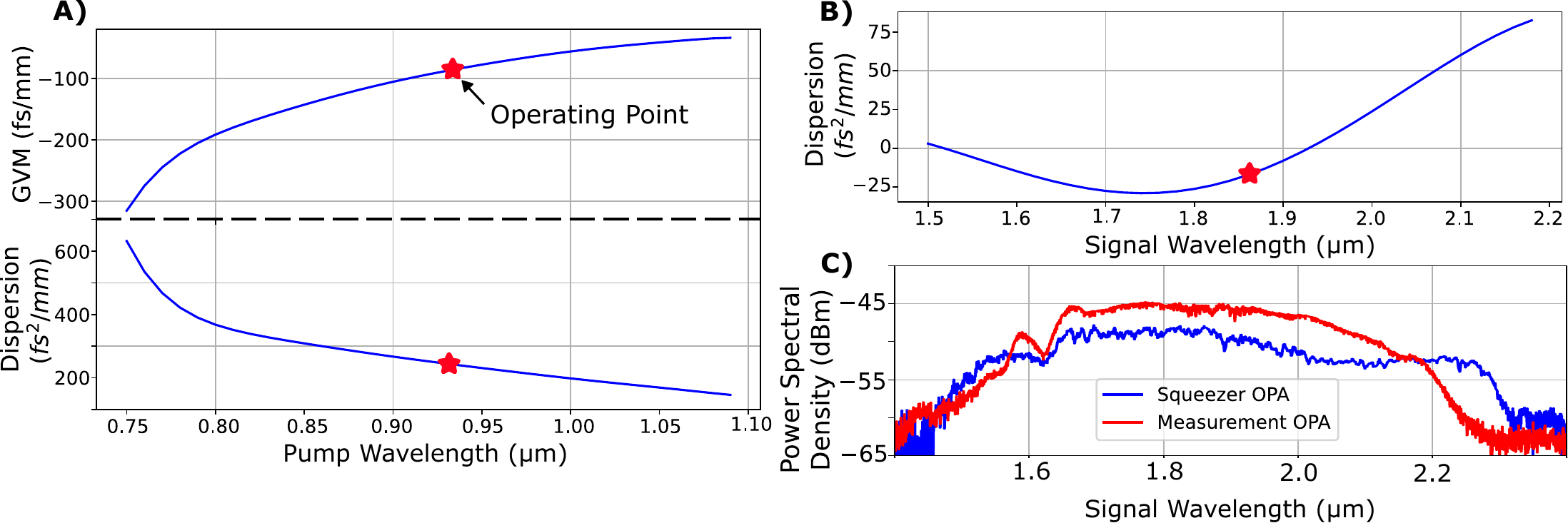}
\caption{\label{fig:specdisp} A) Pump dispersion and group-velocity mismatch (calculated at degeneracy and plotted at the pump wavelength) for our OPA. B) Signal dispersion vs wavelength. C) Parametric generation (vacuum amplification) spectra of our measurement and squeezer OPAs.}
\end{figure*}
\subsection{\label{sec:mmode}Quadrature Measurement}

Figure \ref{fig:fig_1} depicts the scheme used to measure the Wigner function of an SV state encoded in a femtosecond pulse. When the SV enters the nanophotonic amplifier, its in-phase quadrature $\hat{x}_{\phi}$ is amplified by the pump while its out-of-phase quadrature $\hat{p}_{\phi}$ is deamplified. These quadratures map to the original operators of the state as \cite{kalash_wigner_2023}:

\begin{eqnarray}
\label{eq:xpphases}
    \hat{x}_{\phi} = \hat{x}cos(\phi) + \hat{p}sin(\phi) 
    \\
    \hat{p}_{\phi} = \hat{p}cos(\phi) - \hat{x}sin(\phi)
\end{eqnarray}

where $\phi$ represents a relative phase between the pump and amplified signal induced by adjusting the time-delay ($\Delta t$) of the pump. After amplification, the output quadratures can be expressed as: 
\begin{eqnarray}
    \hat{X_{\phi}} = \hat{x}_{\phi} e^{g}
    \\
    \hat{P_{\phi}} = \hat{p}_{\phi} e^{-g}
\end{eqnarray}

where $e^{g}$, the gain of the amplifier, can be controlled by changing the pump power. The operator for the number of photons in the signal field at the output is then:

\begin{eqnarray}
    \hat{N_{\phi}} = \hat{X}_{\phi}^2 + \hat{P}_{\phi}^2 - \frac{1}{2}
    \label{eqn:photonnumber}
\end{eqnarray}

For a large gain, the second and third terms of Eq.\ref{eqn:photonnumber} can be ignored such that:

\begin{eqnarray}
    \hat{N_{\phi}} \approx \hat{X}_{\phi}^2
\end{eqnarray}

By measuring the signal intensity at the output of the OPA for each pulse, we can recover the marginal distribution $P(N,\phi)$, or the probability of detecting $N$ photons at a measurement pump phase $\phi$. 

Because our OPA exhibits low dispersion over a broad bandwidth while operating in the type-0 phase matching configuration, it amplifies many orthogonal spectro-temporal modes simultaneously \cite{u2006generation}, all of which contribute to the measured $P(N,\phi)$. These modes can be calculated analytically by computing the joint spectral intensity (JSI) function of the signal and idler. Our JSI and the corresponding modes are computed and plotted in Appendices using the dispersion parameters calculated from the waveguide geometry and the Heisenberg propagators derived in \cite{houde2023waveguided}. Because our JSI is inseparable ($JSI(\omega_s, \omega_i) \neq \phi(\omega_s) \phi(\omega_i)$), it is composed of multiple independent modes $\Phi(\omega_s, \omega_i)$ of the form $\Phi(\omega_s, \omega_i) = \phi(\omega_s) \phi(\omega_i)$.

At the photodetector, we use a combination of a 1700-nm long-pass filter and a 1950-nm short-pass filter to suppress contributions from higher-order modes. Using the measured $P(N)$ for vacuum amplification (i.e. when no signal field is sent into the OPA), we calculate a Schmidt number of 1.35 modes after filtering from the definition of $g^{(2)}$ in Ref. \cite{florez2020pump}. Contributions from the higher-order modes are quantified and separated by fitting a two-mode photon number distribution to determine the gain of the fundamental mode as a function of $\phi$. As our Schmidt number of 1.35 is close to single-mode (i.e. 1), we limit our fit to the first two modes as contributions from higher-order modes are relatively small. For a single mode, the photon number distribution is

\begin{equation}
    P^{[1]}_{\langle N \rangle}(N, \phi) = \frac{1}{\sqrt{2\pi N\langle N(\phi) \rangle}} e^{-\frac{N}{2\langle N(\phi) \rangle}}
\end{equation}

where $\langle N(\phi) \rangle$ is the squeezing-dependent average photon number. Convolving this distribution with itself yields the two-mode distribution \cite{lordi2024quantum}:

\begin{equation}
\label{eq:2mode}
    P^{[2]}(N, \phi) = \int_0^{\infty} P^{[1]}_{\langle N_1 \rangle}(N-n,\phi)P^{[1]}_{\langle N_2 \rangle}(n,\phi) dn
\end{equation}

This convolution is the probability of detecting a total of $N$ photons across both modes as the photodetector cannot distinguish between photons arriving from different modes. These fitted distributions are then sampled and used as inputs to a maximum-likelihood reconstruction algorithm to recover the density matrix and Wigner function \cite{bidleoptical}.

\subsection{\label{sec:dispeng}Dispersion Engineering}

\begin{figure*}[t]
\includegraphics[width=\textwidth]{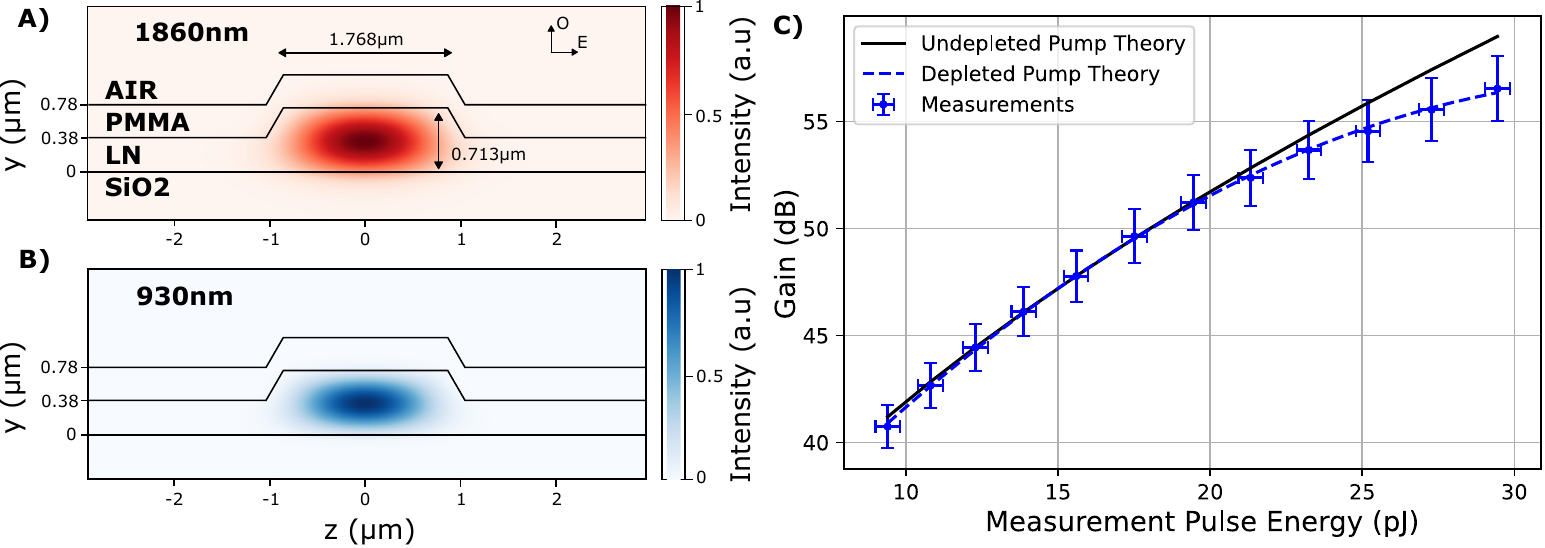}
\caption{\label{fig:mode_figure} A) Signal spatial mode within our waveguide. Text on the right denotes the material stack-up with air, poly methyl methacrylate (PMMA), lithium niobate (LN) and silicon dioxide (SiO2). Dimensions are indicated by the vertical and horizontal measurements. The ordinary and extraordinary crystal axes are denoted at the top right. B) Pump spatial mode. C) OPA gain vs measurement pump energy. Error bars are calculated from pump and signal coupling stability measurements taken before gain measurements. The depleted pump theory is taken from the Gaussian limit defined in \cite{chinni2024beyond}.}
\end{figure*}

A key advantage of using nanophotonics is the ability to control the dispersive properties of the waveguides used to realize photonic circuits. Figure \ref{fig:mode_figure} depicts our ridge waveguide geometry and material stack used for our OPAs. We fabricate our OPAs on a thin-film lithium niobate (TFLN) on silica on silicon wafers available from NanoLN. By adjusting the width, etch depth, and thin-film thickness of the waveguide, we can manipulate its dispersive properties to achieve low group velocity dispersion (GVD) at the pump and signal wavelengths while simultaneously minimizing the walk-off from group velocity mismatch (GVM) between the pump and signal \cite{ledezma2022intense}. Figures \ref{fig:specdisp}A-B plot the GVD for the signal/pump and GVM at degeneracy for the waveguide geometry in Fig. \ref{fig:mode_figure}A. The operating wavelengths are marked with a red star. This regime of operation allows for our OPA to exhibit high gain over a broad bandwidth, making it an ideal tool for studying quantum states encoded in ultrafast pulses. Our waveguide geometry achieves a GVD of $-17.3~\mathrm{fs^2/mm}$ at our signal wavelength of 1860 nm, $\mathrm{244~fs^2/mm}$ at our pump wavelength of 930 nm, and a GVM of $\mathrm{-87~fs/mm}$. By comparison, bulk lithium niobate has a GVD of $\mathrm{13.3~fs^2/mm}$ at 1860 nm, a GVD of $\mathrm{341~fs^2/mm}$ at 930 nm, and a GVM of $\mathrm{203~fs/mm}$. A thin layer of poly-methyl methacrylate (PMMA) is deposited via spin-coating on top of the waveguide that acts to perturb the effective refractive index of the waveguide and offer some tunability of the phase matching condition during fabrication. We note that our design is not perfectly optimal, and that geometries with lower dispersion have been demonstrated \cite{ledezma2022intense}. Ideally, the dispersion of the squeezer and measurement OPAs should be chosen such that the overlap of their fundamental modes is maximized while also minimizing the GVM and GVD. These goals are not always mutually compatible, and hence there is some trade-off. For this work, we focus on minimizing GVM and GVD while relying on filtering to enforce mode overlap.

Figure \ref{fig:specdisp}C shows the parametric generation (i.e. vacuum amplification) spectra for the 2.5-mm OPA used for SV generation and the 5-mm OPA used for measurement. These spectra are taken at a pump pulse energy of 30 pJ. Both OPAs exhibit a 3-dB gain over $>$ 20 THz as a result of low GVD at the signal wavelength and signal gains $>$ 40 dB thanks to low GVM. Compared to the squeezer, the measurement OPA has a slightly narrower and more intense gain spectrum as the extra 2.5-mm of propagation allows for greater walk-off far from degeneracy and more interaction time near degeneracy. 

Operating with low dispersion allows us to utilize ultrafast femtosecond pulses at both the pump and signal frequencies. Shorter pulses are advantageous for nonlinear processes such as OPA because the high peak power from temporal confinement can achieve a stronger nonlinear interaction for the same pulse energy. This interaction is also enhanced by the tight spatial mode confinement offered by ridge waveguides. Furthermore, by shortening the length of the pump and signal pulses in the time domain, subsequent pulses can be packed closer together in time, enabling a significant boost in clock speeds for time-multiplexing. This boost also allows larger quantum circuits to be realized in cluster state architectures \cite{asavanant2019generation, larsen2019deterministic, madsen2022quantum}. Based on the dispersive properties of our 5-mm measurement OPA and our 70-fs pump source, we estimate an upper bound on the temporal length of our SV of 154.3 fs, which gives a theoretical maximum of 6.5 THz for our clock rate \cite{wasilewski2006pulsed}. This clock rate can be increased further by using a shorter OPA and a higher pump power, or more advanced dispersion engineering.

\subsection{\label{sec:pumpdep}Pump Depletion}

An important consideration for OPAs as quantum measurement devices is the depletion experienced by the pump during amplification. As the pump energy increases, the nonlinear interaction becomes more efficient, causing the pump pulse to deplete while propagating in the OPA and the gain to saturate. Figure \ref{fig:mode_figure}C shows the measured on-chip OPA gain vs the input pump energy. The black line represents the theoretical undepleted pump gain $G_{dB} = 10\log_{10}(e^{b\sqrt{E}})$ where $E$ is the pump energy in Joules and $b$ is a found by fitting the pump energy vs gain curve in the undepleted regime. The dashed blue line represents the theoretical gain when accounting for pump depletion up to second order in the Gaussian limit \cite{chinni2024beyond}. Around 20 pJ, the measured gain begins to saturate relative to the theoretical gain due to pump depletion effects.

Pump depletion impacts our measurement in two ways. First, the larger photon number components of the state being measured experience less gain than smaller photon number components. This phenomenon, often known as gain saturation for classical amplifiers, suppresses the larger photon number contributions to the measured distributions and distorts our measurement. This distortion can be mitigated by improving output coupling losses to allow operation at a lower pump power. We investigate this distortion experimentally in section \ref{sec:slowsque}. Second, pump depletion can be thought of as a semi-deterministic transfer of energy from the pump pulse to the signal pulse. This causes the peak of the photon number distribution to shift away from $N = 0$. We observe this effect in classical simulations that use a split-step Fourier method to simulate single-mode and multimode vacuum measurements. These simulations are detailed in the supplementary. This effect has also been observed in single-mode quantum simulations \cite{chinni2024beyond} and experimentally \cite{florez2020pump}. We address this issue when fitting our data with 2-mode distributions by fitting a constant offset along the $N$ axis, leading to good agreement between the measured data and fitted distributions.

\section{\label{sec:expsetup} Experimental Setup}

\begin{figure*}[ht!]
\includegraphics[width = \textwidth]{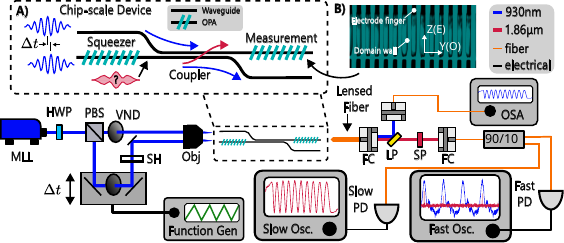}
\caption{\label{fig:exp_setup} Experimental setup. MLL: Titanium:sapphire tunable mode-locked laser. PBS: polarizing beam splitter. VND: variable neutral density filter. SH: mechanical shutter. Obj: reflective objective. FC: reflective fiber collimator. LP and SP: long-pass and short-pass wavelength filters. 90/10: fiber splitter with 90\% going to the fast detector and 10\% going to the slow detector. PD: photodetector. Fast Osc: 80 GSPS 40 GHz oscilloscope. Slow Osc: 100 MSPS 10 MHz oscilloscope.}
\end{figure*}

Figure \ref{fig:exp_setup} shows the experimental setup. An 80~MHz mode-locked titanium-sapphire laser (MLL) sends 70-fs pulses to a half-wave plate (HWP) and polarizing beam splitter (PBS) combination to control the power splitting between the measurement and squeezer beam paths. Pump light for the squeezer OPA is sent to a delay stage used to scan $\phi$ in equation \ref{eq:xpphases}. A shutter is also placed in this path to block the squeezer beam for shot-noise calibration. Two continuously-variable neutral density filters (VNDs) are used in both paths to fine-tune the input power. The paths are combined on a reflective objective (Obj) and focused onto the waveguide inputs for the OPAs. Figure \ref{fig:exp_setup}A shows the layout of the nanophotonic circuit. In the squeezer OPA, the vacuum field around 1860~nm is squeezed by the pump to generate squeezed vacuum (SV). The SV and the remaining pump light enter an adiabatic coupler that passes at least 55\% of the SV and 5\% of the pump into the measurement OPA. Inside the measurement OPA, the SV is amplified by a strong pump pulse (40~pJ) to macroscopic photon levels detectable by a fast photodetector. A lensed fiber collects light from the output of the measurement OPA and directs it to a series of free-space filters used to remove remaining pump light, including the previously mentioned 1700-nm long pass and a 1950-nm short pass for limiting the measured mode number. From here, the filtered SV is collected into single-mode fiber and enters a 90/10 splitter where 90\% is sent to a fast (10 GHz) photodetector and fast (80 GSPS 40 GHz) oscilloscope used to resolve individual pulses while the remaining 10\% is sent to a slow (1 MHz) detector and slow (100 MSPS 10 MHz) oscilloscope used to detect the amplification fringe and determine the measurement phase $\phi$.

\section{Data Analysis}

\subsection{\label{sec:shotnoise}Shot-noise Calibration}
%Example measurement figure here

To calibrate our squeezing and anti-squeezing measurements, we measure the $\phi$-dependent shot noise using the slow (1 MHz) photodetector and an optical spectrum analyzer (OSA) which replaces the fast detector in Fig. \ref{fig:exp_setup}. The coupler in Fig. \ref{fig:exp_setup}A couples some of the 930-nm pump light used to drive the squeezer into the measurement OPA, causing it to interfere with the pump light sent into the upper port of Fig. \ref{fig:exp_setup}A. Because of this interference, the gain inside the measurement OPA, and hence the shot noise, is dependent on $\phi$. 

Data from the shot-noise calibration procedure are show in figure \ref{fig:slow_data}A. During calibration, the mechanical shutter is initially left open. As the relative delay of the squeezer path is scanned, the measurement amplifier oscillates between measuring squeezed (with destructive pump interference) and anti-squeezed (with constructive pump interference) quadratures, resulting in the black curve in Fig. \ref{fig:slow_data}A. The remaining pump light collected from the output is monitored on a second OSA to determine the strength of the pump interference. The shutter is then closed and the baseline shot noise is measured by the OSA resulting in the gray curve in Fig. \ref{fig:slow_data}A. To find the shot-noise level corresponding to destructive pump interference (i.e. the shot-noise minimum), the intensity of the measurement OPA pump is lowered until the measured power at the pump OSA matches the minima of the previously measured pump interference. The shot-noise minimum is then measured on the slow photodetector and OSA resulting in the red curve in Fig. \ref{fig:slow_data}A. This same procedure is repeated for the shot-noise maximum to calibrate the anti-squeezing and obtain the blue curve. The pump OSA is set to measure around 970 nm to minimize any pump depletion effects. After shot-noise calibration, the squeezing is calculated as:

\begin{eqnarray}
\label{eqn:squeezing}
    S_{\pm}^\phi[dB] = 10 \log_{10}\Big[\frac{\braket{N(\phi)}}{\braket{N_{vac}}}\Big]
\end{eqnarray}

where $\braket{N(\phi)}$ is proportional to the current measured at the detector, and $\braket{N_{vac}}$ is the calibrated shot noise. We note that the variance of the measured quadrature is proportional to the average photon number for a state with zero mean field such as SV \cite{nehra2022few}.

\subsection{\label{sec:slowsque}Slow Photodetector Measurements}

\begin{figure*}[t]
\includegraphics[width = \textwidth]{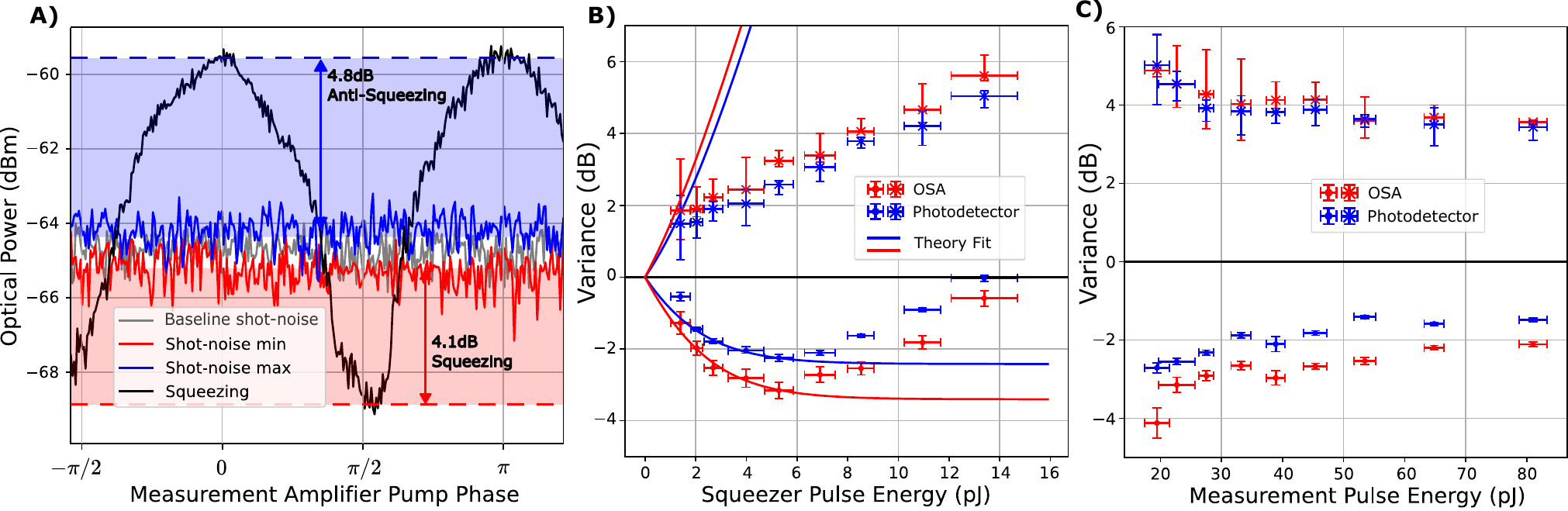}
\caption{\label{fig:slow_data} A) Squeezing measurement taken on the OSA. The black curve is the amplification fringe produced from squeezed vacuum amplified in the measurement OPA. The grey curve is the baseline shot noise at the pump power used to measure the black curve. The red and blue curves are the corrected shot-noise levels to account for pump interference in the measurement OPA. B) Variance relative to vacuum vs squeezer pulse energy with a measurement pulse energy of 30 pJ. C) Variance relative to vacuum vs measurement pulse energy. Error bars are calculated from shot-noise variations during each measurement.}
\end{figure*}

Figure \ref{fig:slow_data}B shows the measured squeezing on the slow photodetector and OSA as a function of the input pulse energy for the squeezer waveguide with a measurement pulse energy of 30 pJ. For these measurements, the 1950-nm long-pass filter is removed to accentuate the impact of multiple modes (a Schmidt number of 2.8). Because the OSA uses a 2-nm bandpass filter, it can better suppress higher-order modes to isolate the squeezing in the fundamental mode \cite{nehra2022few, houde2023waveguided}. For the photodetector, the presence of higher-order modes with less squeezing reduces overall measured squeezing as the photodetector is not mode-selective and measures the squeezing as averaged over all modes. 

At low squeezer pulse energies, the measured squeezing is limited by the gain available in the squeezer OPA and grows as the pulse energy is increased. After 5 pJ, the observed squeezing begins to decrease with increasing pulse energy. Two phenomena contribute to this behavior. First, the finite phase noise of the pump laser causes a portion of the anti-squeezed quadrature to leak into the measurement of the squeezed quadrature. Raising the squeezer gain increases the contribution of the anti-squeezing to the measurement and reduces the measured squeezing. This increased gain produces more squeezing, but the noise reduction in the squeezed quadrature is asymptotically limited by the finite efficiency of the coupler while the noise of the anti-squeezed quadrature is not (see solid theory curves in Fig. \ref{fig:slow_data}B). Second, gain saturation in the measurement amplifier increases with input power and further degrades measured squeezing. 

The effects of gain saturation on measured squeezing can be more easily seen in Fig. \ref{fig:slow_data}C which plots the squeezing values vs the measurement pulse energy for a fixed squeezer energy of 5 pJ. As the measurement pulse energy is decreased, the behavior of the measurement OPA becomes more linear with less gain compression, leading to higher measured squeezing and anti-squeezing. In the undepleted pump regime, measured squeezing and anti-squeezing will follow Eq.\ref{eq:squeezing_loss} and remain unchanged with small changes in pump energy. Hence, we know all measurements in Fig. \ref{fig:slow_data}B are taken in the pump depletion regime as the measured anti-squeezing deviates from the theoretical model calculated using the efficiency of the coupler. This is also confirmed from Fig. \ref{fig:mode_figure}C where pump depletion effects take hold around 20 pJ. Below 20 pJ, the shot noise no longer clears the electronic noise floor of the OSA, preventing squeezing measurements at lower measurement pump energies. For this reason, data points in Fig. \ref{fig:slow_data}B are taken at a pump energy of 30pJ to improve SNR and shrink vertical error bars while sacrificing some measurable squeezing (3 dB at 30 pJ pump compared to 4.1 dB at 20 pJ pump in Fig. \ref{fig:slow_data}B and C respectively).   

The solid lines plotted in Fig. \ref{fig:slow_data}B are calculated from the theoretical measured squeezing as a function of the coupling loss from the squeezer to measurement OPA. As SV passes through the coupler, some of the light is lost as a result of fabrication imperfections in the coupler. This light is replaced with vacuum noise, causing lower squeezing and anti-squeezing values to be measured \cite{henry1996quantum}. For a measurement efficiency $\eta$, the measured squeezing can be modeled as:

\begin{eqnarray}
\label{eq:squeezing_loss}
    S_{\pm}^\eta[dB] = 10\log_{10} [(1-\eta) + \eta e^{\pm 2r}]
\end{eqnarray}

where $r$ is the on-chip squeezer gain in natural log units. Using this fit, we estimate an effective $\eta$ of 55\% at 1860 nm. This fit includes squeezing lost to inefficiencies in our coupler as well as losses to nonlinear behavior in the measurement OPA. We expect $\eta$ to be dominated by the coupler losses given the characterization measurements taken previously of this device \cite{nehra2022few}. The vacuum noise injected from losses contaminates the SV, resulting in a state with reduced squeezing and anti-squeezing measured by the measurement OPA. 

Equation \ref{eq:squeezing_loss} also allows us to calculate the expected anti-squeezing, however, the measured data have a stark divergence from the expected trend. This is a result of gain-compression in the measurement OPA. Because the measurement OPA operates in the pump-depleted regime, the anti-squeezed vacuum cannot be sufficiently amplified to reflect the true anti-squeezing entering the OPA. The same is true for the squeezing measurements as the shot-noise level becomes compressed, leading to lower measured squeezing.

\subsection{\label{sec:fastm}Pulse-to-pulse Measurements}

\begin{figure*}
    \includegraphics[width=\textwidth]{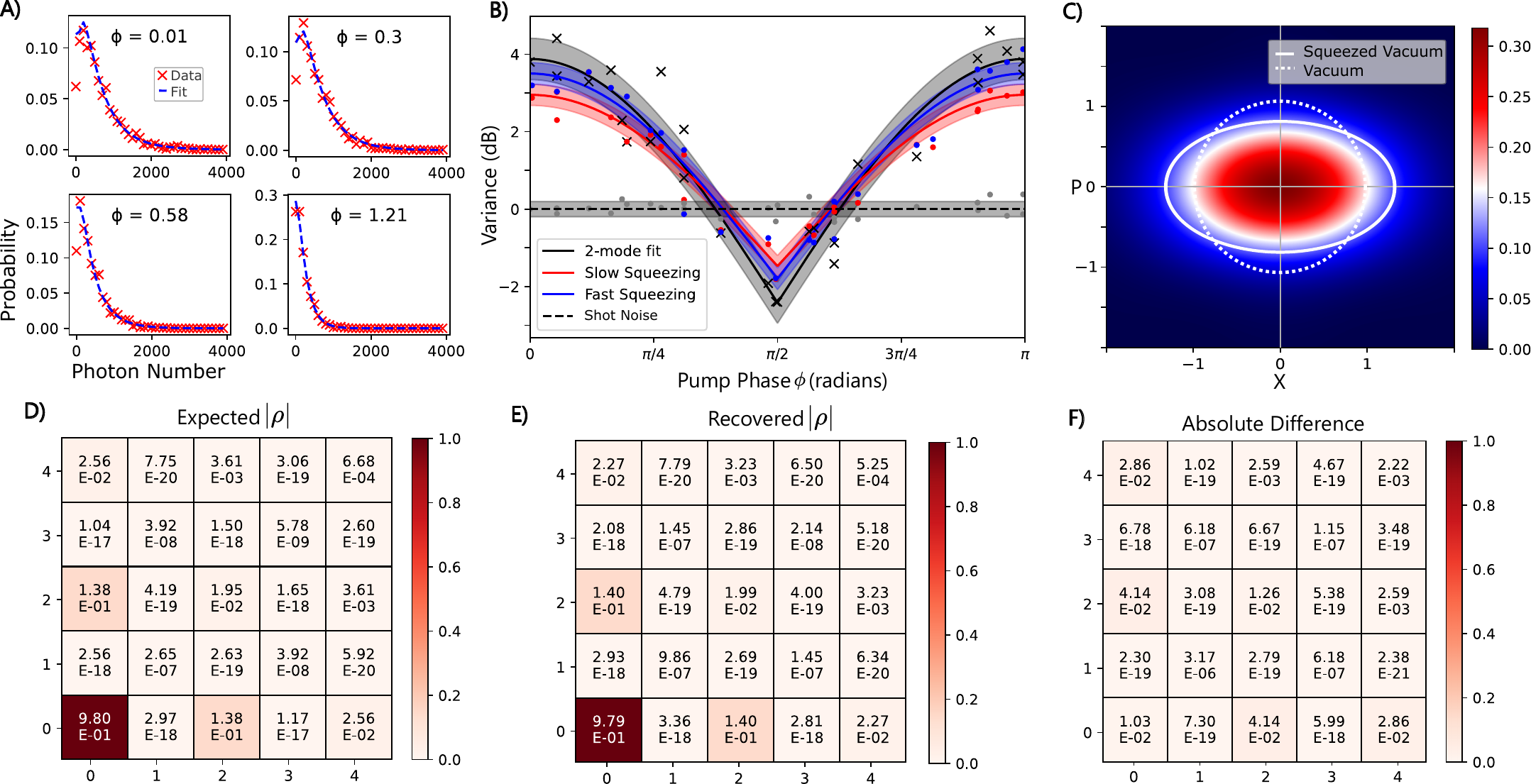}
    \caption{\label{fig:fast_data} A) Sample photon number distributions from fast measurements at different $\phi$. Red points are data, and the blue curve is a two-mode fit. B) Photon number variance vs $\phi$ for slow, fast (pulse-to-pulse) measurements and the first mode of a two-mode fit. Solid lines are the expected cosine dependence taken from Ref.\cite{kalash_wigner_2023}. C) Recovered squeezed vacuum Wigner function. Solid and dashed while lines lie along the $1/(e \pi)$ contours. D) Expected density matrix ($\rho$) based on measured squeezing and anti-squeezing. E) Recovered $\rho$. F) Absolute difference between expected and recovered $\rho$.}
\end{figure*}

Fast photodetector measurements resolve the pulse-to-pulse intensity and provide the statistical information necessary to resolve $P(N, \phi)$ and recover the Wigner function. During a fast measurement, the shutter is initially left open, and the fast oscilloscope is triggered using the voltage applied to the piezo driving the squeezer's delay stage to coincide with the middle of the $\phi$ scan. Data from the slow oscilloscope are used to determine $\phi$ and estimate the coupling stability based on measurement-to-measurement variations in the shot noise. After recording data, the shutter is closed to block the squeezer and the measurement is repeated to find the calibrated shot-noise level. 

Figure \ref{fig:fast_data} shows results gathered from fast photodetector measurements for a squeezer pulse energy of 5 pJ and a measurement pulse energy of 45 pJ. This measurement pulse energy was chosen to satisfy SNR requirements at the fast photodetector given the loss between the chip and detector. Sample measured histograms at four different phases are plotted in Fig. \ref{fig:fast_data}A. Each histogram corresponds to a single data point in Fig. \ref{fig:fast_data}B. For each measurement, we fit a two-mode photon number distribution model to extract the variance of the fundamental mode as a function of $\phi$. Figure \ref{fig:fast_data}B shows these measurements for three cases: slow photodetector measurements, fast photodetector measurements, and the two-mode fit. Slow measurement values are calculated from the slow oscilloscope trace using the method outlined in section \ref{sec:slowsque}. Because of the photon number offset resulting from pump depletion, the numerator and denominator of Eq.\ref{eqn:squeezing} have a small but constant offset, resulting in the measured squeezing and anti-squeezing being underestimated. Fast measurement values are calculated using the variance of the photon numbers measured at the fast detector. This overcomes the offset problem and measures more squeezing and anti-squeezing, but these measurements are still contaminated with remaining contributions from higher-order modes. The two-mode fit addresses this problem by isolating the variance as a function of $\phi$ for the fundamental mode, leading to higher measured squeezing and anti-squeezing. After performing this fit for all measurements, the distributions are re-sampled and fed into a maximum likelihood algorithm that calculates the Wigner function (Fig. \ref{fig:fast_data}C). We extract a squeezing in the fundamental mode of 2.41 $\pm$ 0.34 dB and an anti-squeezing of 3.87 $\pm$ 0.61 dB. Figure \ref{fig:fast_data}E shows the recovered density matrix corresponding to the Wigner function plotted in Fig. \ref{fig:fast_data}C. We show the expected density matrix given the measured squeezing and anti-squeezing in Fig. \ref{fig:fast_data}D, and the absolute difference between the expected and recovered in Fig. \ref{fig:fast_data}F where we calculate a fidelity of 0.9998 $\pm$ 0.0001. This fidelity serves as a sanity-check to ensure our data analysis is not introducing distortions into the measurement. While our fitting strategy limits us to states with known distributions, we can overcome this limitation with tighter bandpass filtering to better reject higher-order modes. Our losses between the chip and fiber (10 dB) and from our filter setup (6 dB) currently prevent us from tighter filtering as the remaining light is no longer intense enough to provide a sufficient signal-to-noise ratio at the fast photodetector.

\begin{figure*}[ht!]
    \includegraphics[width=\textwidth]{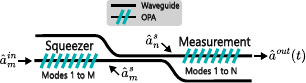}
    \caption{\label{fig:circuit} A diagram of the squeezer and measurement OPA circuit. }
\end{figure*}

\section{Conclusion}
We have demonstrated OPA-based Wigner tomography for squeezed vacuum in the ultrafast regime using dispersion-engineered TFLN. This is, to the best of the authors' knowledge, the first demonstration of all-optical quantum state tomography on-chip. The low dispersion of our OPA design allows operation with ultra-short pulses to exceed the bandwidth of traditional measurement techniques and access a new regime of THz repetition-rate measurement and computation. Achieving THz clock rates is a fundamental advancement for time-multiplexed systems such as cluster states and measurement-based quantum computation as higher clock rates translate to faster computation and allow for larger quantum circuits to be realized.

Our THz claim is based on dispersion calculations derived from the waveguide geometry measured using atomic force microscopy. We also note that similar dispersion-engineered OPAs with faster clock rates than the one presented here have been experimentally characterized \cite{guo2022femtojoule}. Speed limitations imposed by the photodetector bandwidth are present in our experiment, but can be overcome with fast demultiplexing schemes which we elaborate in Appendices.

Because we only measure intensity, our current experimental implementation is limited to tomography of states with zero mean field. However, OPA-based state tomography can be generalized to any arbitrary state by implementing either homodyne detection or a displacement scheme at the output, both of which are mature technologies \cite{nehra2024all, nehra2019state, laiho2010probing, larsen2021deterministic, kawasaki2024broadband}. As discussed in earlier sections, experimental limitations stemming from high losses between the chip and fast photodetector introduce measurement distortion from pump depletion and multimode effects. Both of these problems can be solved with narrower bandpass filtering and modest improvements in detection efficiency to allow for the measurement OPA to operate in the undepleted pump regime. 

To conclude, we show that dispersion-engineered nanophotonic OPAs can serve as quantum measurement devices, paving the way for ultrafast quantum information processing to be realized in a room-temperature chip-scale platform.

\section{Acknowledgments}

Device nanofabrication was performed at the Kavli Nanoscience Institute (KNI) at Caltech. The authors gratefully acknowledge support from ARO grant no. W911NF-23-1-0048, NSF grant no. 1918549, AFOSR award FA9550-23-1-0755, DARPA award D23AP00158, the Center for Sensing to Intelligence at Caltech, the Alfred P. Sloan Foundation, and NASA/JPL.

\section{Data Availability Statement}

The data that support the findings of this article are not publicly available upon publication because it is not technically feasible and/or the cost of preparing, depositing, and hosting the data would be prohibitive within the terms of this research project. The data are available from the authors upon reasonable request.

\begin{figure*}[t]
    \includegraphics[width=\textwidth]{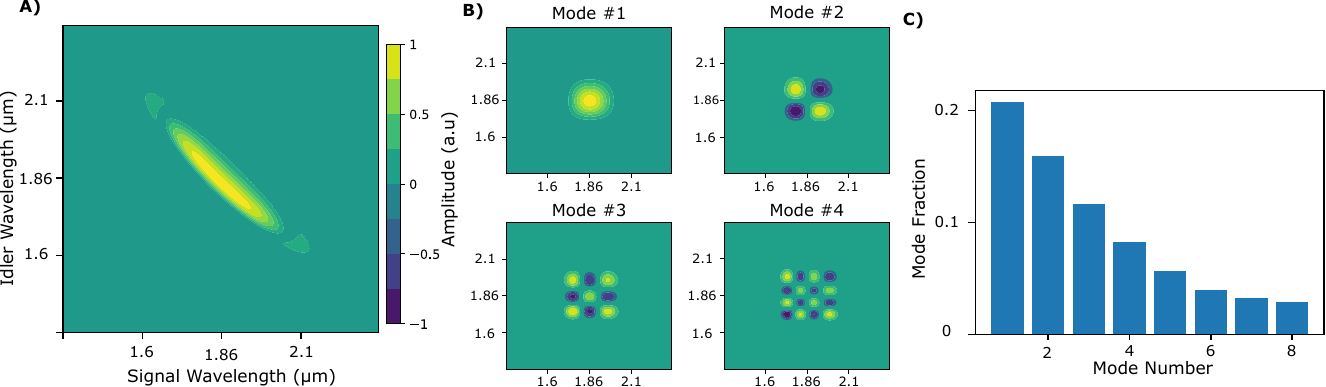}
    \caption{\label{fig:jsi_fig} A) A plot of the joint-spectral intensity function of the squeezed vacuum produced by our OPA. B) The first four modes of the JSI calculated via the Bloch-Messiah decomposition. C) Plot of the mode number vs proportion of mode present in the JSI.}
\end{figure*}

\section{Appendix 1: Multimode Cascaded Squeezing in the Heisenberg picture}

We use the formalism of Ref.\cite{wasilewski2006pulsed} to derive the output of two cascaded multimode OPAs: a squeezer OPA and a measurement OPA. Figure \ref{fig:circuit} shows the circuit diagram. We begin with the state after the squeezer OPA (denoted with superscript $s$) and decompose the annihilation operator into the squeezer OPA's eigenmodes (denoted by subscript $m$):
\begin{align}
    \hat{a}^s(t) &= \sum_m \psi^{s*}_m(t) \hat{a}^s_m \\
    &= \sum_m \psi^{s*}_m(t) (\cosh r^s_m \hat{a}^{in}_m + \sinh r^s_m \hat{a}^{in\dag}_m) \nonumber
\end{align}
where $r_m^s$ is the squeezing parameter of mode $m$. Each eigenmode of the squeezer OPA is independently squeezed. To simplify later expressions, we assume that $r_m^s = r_m^s(\phi)$ where $\phi$ is the phase of the squeezed vacuum relative to the measurement pump. In our experiment, we inject vacuum into the squeezer OPA, and so $\hat{a}^{in} \equiv \hat{a}^{vac}$.

After the squeezer OPA, the state is placed into the measurement OPA (superscript $ms$ and mode subscript $n$) and the resulting output operator is labeled $\hat{a}^{out}(t)$. The measurement OPA amplifies its own modes as independent single-mode OPAs:
\begin{align}
    \hat{a}^{out}(t) &= \sum_n \psi^{ms*}_n(t)\hat{a}^{out}_n \\
    &= \sum_n \psi^{ms*}_n(t) (\cosh r^{ms}_n \hat{a}^s_n + \sinh r^{ms}_n \hat{a}^{s\dag}_n)  \nonumber
\end{align}
We can write the squeezed operators in the measurement OPA basis as a function of the mode overlaps between the measurement OPA's and squeezer OPA's eigenmodes $\sigma_{mn} = \int d t~ \psi^{ms}_n(t) \psi^{s*}_m(t)$ :

\begin{align}
    \hat{a}^s_n &= \int_{-\infty}^\infty dt~\psi^{ms}_n(t) \hat{a}^s(t) \\
    &=\int_{-\infty}^\infty  dt~\psi^{ms}_n(t) \sum_m \psi^{s*}_m(t) \hat{a}^s_m  \nonumber \\
    &= \sum_m \sigma_{mn} \hat{a}^s_m.  \nonumber
\end{align}

The output annihilation operator is then:
\begin{align}
    \hat{a}^{out}(t) &= \sum_{mn}\psi^{ms*}_n(t)(\cosh r^{ms}_n \sigma_{mn} \hat{a}^s_m + \sinh r^{ms}_n \sigma^*_{mn} \hat{a}^{s\dag}_m) \\
    &= \sum_{mn}\psi^{ms*}_n(t) \nonumber \\
     &(\cosh r^{ms}_n \sigma_{mn} (\cosh r^s_m \hat{a}^{in}_m + \sinh r^s_m \hat{a}^{in\dag}_m)  \nonumber \\
    &~~~~~~~~ + \sinh r^{ms}_n \sigma^*_{mn} (\cosh r^s_m \hat{a}^{in\dag}_m + \sinh r^s_m \hat{a}^{in}_m))  \nonumber \\
    % &= \sum_{mn}\psi^{ms*}_n(t) \int \dd t' \big[\cosh r^{ms}_n \sigma_{mn} (\cosh r^s_m \psi^s_m(t') \hat{a}^{in}(t') + \sinh r^s_m \psi^{s*}_m(t') \hat{a}^{in\dag}(t')) \\
    % &~~~~~~~~ + \sinh r^{ms}_n \sigma^*_{mn} (\cosh r^s_m \psi^{s*}_m(t') \hat{a}^{in\dag}(t') + \sinh r^s_m \psi^s_m(t') \hat{a}^{in}(t'))\big] 
    \nonumber
\end{align}
For large measurement OPA gains, we use the approximation $\cosh r^{ms}_n \approx \sinh r^{ms}_n \approx \frac{1}{2} \exp(r^{ms}_n)$, when $e^{r^{ms}} \gg 1 \gg e^{-r^{ms}}$. Then, the output operator becomes:
\begin{align}
    \hat{a}^{out}(t) = \sum_{mn} \frac{e^{r^{ms}_n}}{2} \psi^{ms*}_n(t) \nonumber &\\ (\sigma_{mn}(\cosh r^s_m \hat{a}^{in}_m + \sinh r^s_m \hat{a}^{in\dag}_m) + \nonumber \\ \sigma^*_{mn}(\cosh r^s_m \hat{a}^{in\dag}_m   + \sinh r^s_m \hat{a}^{in}_m)) 
\end{align}

We can write the photon number operator $\hat{N}$ for an individual pulse as: 
\begin{align}
    \hat{N} &= \int_{-\Delta t}^{\Delta t}dt~ \hat{a}^{out\dag}(t) \hat{a}^{out}(t) \\
    &= \int dt \sum_{m,m',n,n'} \frac{e^{r^{ms}_n + r^{ms}_{n'}}}{4} \nonumber \\& \Big(\psi^{ms}_n(t) (\sigma^*_{mn} (\cosh r^s_m \hat{a}^{in\dag}_m + \sinh r^s_m \hat{a}^{in}_m)  \nonumber \\
    &~~~~~~~ +  \sigma_{mn} (\cosh r^s_m \hat{a}^{in}_m + \sinh r^s_m \hat{a}^{in\dag}_m))\Big)  \nonumber \\
    &~~~~ \Big(\psi^{ms*}_{n'}(t)(\sigma_{m'n'} (\cosh r^s_{m'} \hat{a}^{in}_{m'} + \sinh r^s_{m'} \hat{a}^{in\dag}_{m'})  \nonumber \\
    &~~~~~~~ + \sigma^*_{m'n'} (\cosh r^s_{m'} \hat{a}^{in\dag}_{m'} + \sinh r^s_{m'} \hat{a}^{in}_{m'}))\Big)  \nonumber \\
    &= \sum_{m,n} e^{2 r^{ms}_n} (\Re[\sigma_{mn}] e^{r^s_m} \hat{X}^{in}_m - \Im[\sigma_{mn}]e^{-r^s_m}\hat{P}^{in}_m)^2  \nonumber
\end{align}
where the integral over $\Delta t$ captures the total time-duration of the pulse, and we have used the orthonormality of modes ($\int dt~\psi_n(t) \psi^*_{n'}(t) = \delta_{nn'}$) to insert $\delta_{nn'}$ and $\delta_{mm'}$ and resolve $m^\prime$ and $n^\prime$ in the summation.

%If we have that $\Im[\sigma_{mn}] = 0$, (when the phases of the modes are almost identical, as in the biphoton case (cite Lvovsky) or if the modes have similar shapes) then we have

Bandpass filtering around degeneracy suppresses the imaginary component of $\sigma_{mn}$ as imaginary contributions arise from phase walk-off in the JSI away from degeneracy. Taking $|\Im[\sigma_{mn}]e^{-r_m^s}| << |\Re[\sigma_{mn}]e^{r_m^s}|$, we have: 

\begin{figure*}[t]
    \includegraphics[width=\textwidth]{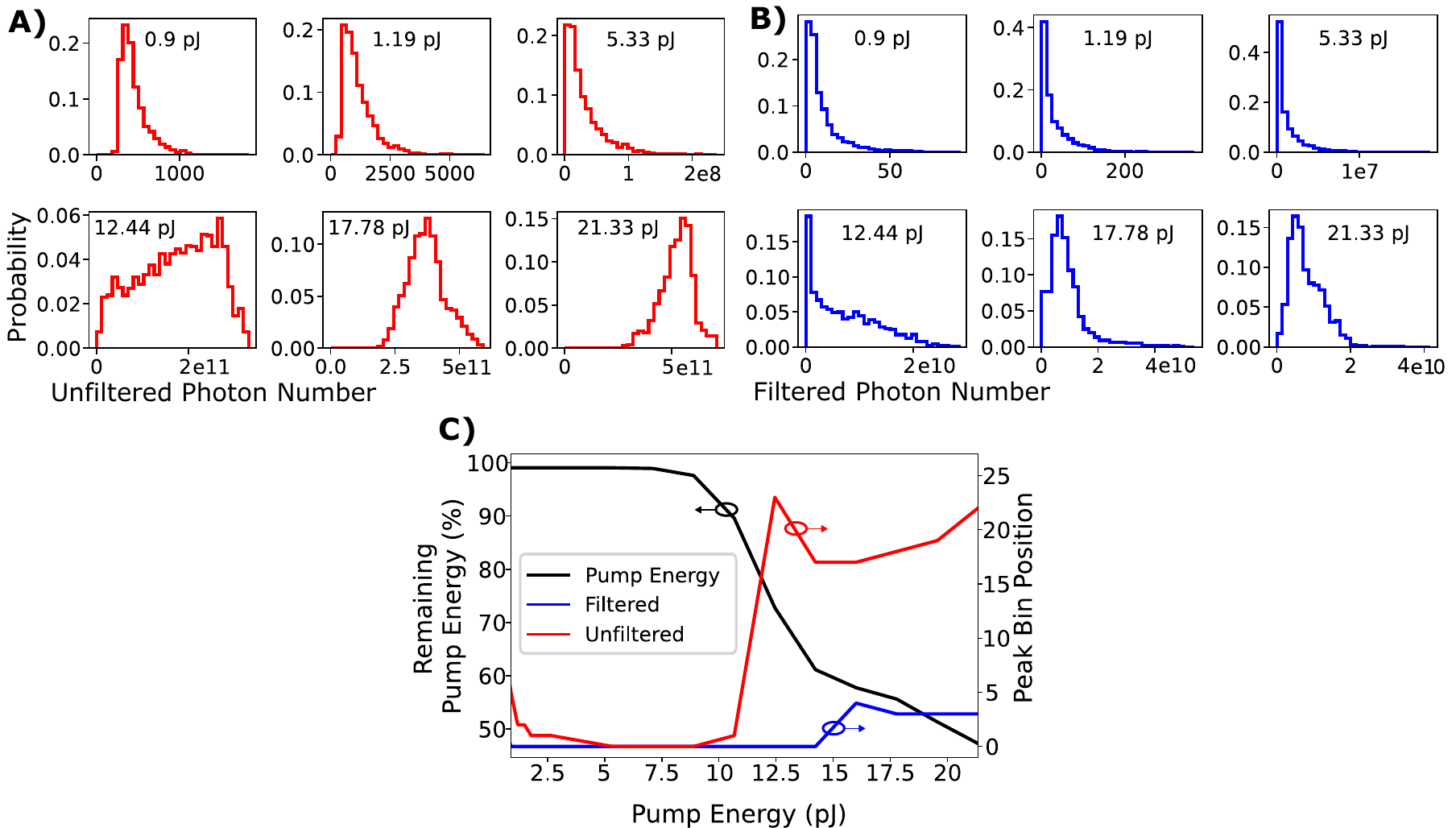}
    \caption{\label{fig:pump} A) Simulated photon number distributions for unfiltered detection across different pump energies. B) Simulated photon number distributions for 20-nm bandpass filtered detection across different pump energies. C) A plot of the remaining pump energy after pulse propagation (lower remaining energy is a more depleted pump), and the peak histogram bin for the unfiltered and filtered simulations.}
\end{figure*}

\begin{align}
\label{eq:n_hat_total}
    \hat{N} = \sum_{m, n} \frac{e^{2 r^{ms}_n + 2 r^s_m}}{4} \sigma_{mn}(\hat{a}^{in}_m + \hat{a}^{in \dag}_m)^2,
\end{align}
If we compare to the case of no squeezing from the squeezer OPA ($\sigma_{mn} = \delta_{mn}, r^s_m = 0 ~\forall~m$):
\begin{equation}
    \hat{N} = \sum_n \frac{e^{2r^{ms}_n}}{4} (\hat{a}^{in}_n + \hat{a}^{in \dagger}_n)^2
\end{equation}
and assume $M$ squeezer modes and $N$ measurement modes, we show that there are now $MN$ modes with effective squeezing parameters $r^{eff}_{m,n} = r^{ms}_n + r^s_m$, whose contributions are weighted by $\sigma_{mn}$. From here, we can return the quadrature representation $\hat{X} = \frac{1}{\sqrt{2}}(\hat{a} + \hat{a}^\dagger)$ to simplify Eq.\ref{eq:n_hat_total} as:

\begin{equation}
\label{eq:n_hat_x}
    \hat{N} = \sum_{m, n} \frac{e^{2 r^{ms}_n + 2 r^s_m}}{8} \sigma_{mn}(\hat{X}_{m}^{in})^2
\end{equation}

In our experiment, we make the assumption that only the first two modes provide significant contributions to the measured statistics as our Schmidt number after filtering is 1.35, and so we truncate $m$ and $n$ to the range [1,2]. Furthermore, we know that $\sigma_{mn}$ where $m$ and $n$ are of different parity are zero. This simplifies Eq.\ref{eq:n_hat_x} to:

\begin{equation}
      \hat{N} = \frac{e^{2 r^{ms}_1 + 2 r^s_1}}{8} \sigma_{11}(\hat{X}_{1}^{in})^2 + \frac{e^{2 r^{ms}_2 + 2 r^s_2}}{8} \sigma_{22}(\hat{X}_{2}^{in})^2
\end{equation}

This allows us to model the measured photon number distribution as the sum of two independent photon number distributions. During the experiment, we treat the fast photodetector as a macroscopic photon number resolving detector such that $N \propto I_d$ where $I_d$ is the current measured on the detector integrated over a single pulse. For a single mode, the photon number distribution is \cite{kalash_wigner_2023}: 
\begin{equation}
    P^{[1]}_{\langle N \rangle}(N, r) = \frac{1}{\sqrt{2\pi N\langle N \rangle(r)}} e^{-\frac{N}{2\langle N \rangle(r)}}
\end{equation}
with average photon number $\langle N \rangle(r)  = e^{2r}/4$. Since the photon number contributions from each mode are summed at the detector, we can model the total distribution as the convolution of two single-mode distributions:

\begin{equation}
\label{eq:2mode2}
    P^{[2]}(N, r) = \int_0^{\infty} P^{[1]}_{\langle N_1 \rangle}(N-n, r)P^{[1]}_{\langle N_2 \rangle}(n, r) dn
\end{equation}

By fitting Eq.\ref{eq:2mode2} to the measured photon number distributions, we isolate the squeezing parameter of the first mode from higher-order modes.

%What we want is to find the squeezing in each input mode (\{$r^s_m$\}). We assume we have \{$r^{ms}_n$\} and $\sigma_{mn}$. The voltage $V_{pc}$ measured on the photodetector for a certain pulse, which can resolve individual pulses but not the temporal shape within the pulse, is
%\begin{equation}
%    V_{pc} = I_{pc} R = \eta E_{pc} R = N \hbar \omega f_{rep} \eta R
%\end{equation}
%for $I_{pc}$ the photocurrent, $R$ the impedance of the detector, $\eta$ the responsivity of the photodetector, $E_{pc}$ the energy of the light reaching the detector, $N$ the number of photons within the pulse, $\omega$ the frequency of the light in rad/s, and $f_{rep}$ the repetition rate of the laser. We see that the photodetector effectively measures the photon number integrated over the pulse. We are interested in the probability distribution of these pulses $p(V_{pc}) \propto p(N)$.

\section{Appendix 2: Calculation of the Joint Spectral Intensity and its Modes}

Figure \ref{fig:jsi_fig}A shows the JSI of the squeezed vacuum we measure. We calculate our JSI using the dispersive parameters obtained from mode simulations, the input pump pulse energy, and the resources from Ref.\cite{houde2023waveguided}. The modes of the JSI are found by performing the Bloch-Messiah decomposition which calculates the modes such that the correlations between pairs of modes are minimized. The first four modes are plotted in Fig. \ref{fig:jsi_fig}B and the fraction of the first eight modes comprising the JSI is shown in Fig. \ref{fig:jsi_fig}C. The JSI decomposition reveals a Schmidt mode number of 7.1. This same calculation can be repeated for the measurement OPA resulting in a Schmidt number of 3.9. Filtering before detection reduces this number to 1.35.

\begin{figure*}[t]
    \includegraphics[width=\textwidth]{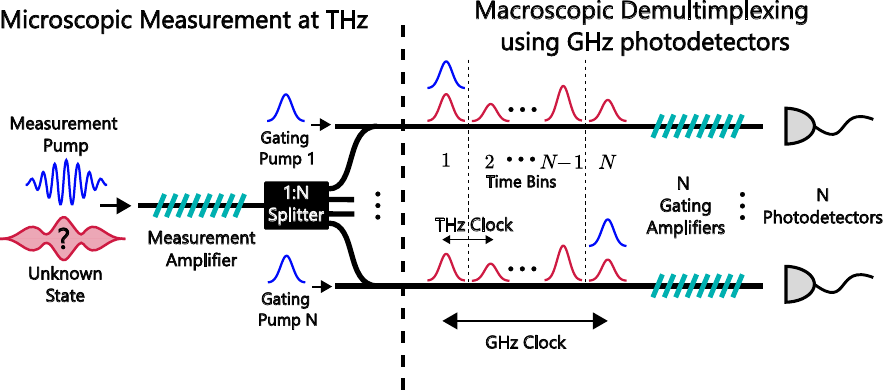}
    \caption{\label{fig:multiplex} A demultiplexing circuit used to transition from a clock rate $f$ to a clock rate $f/N$.}
\end{figure*}

\section{Appendix 3: Pump Depletion in Simulation}

As discussed in the main text, we model the effects of pump depletion as a shift in the peak of the photon number distribution. Figure \ref{fig:pump} shows semi-classical simulation results of vacuum amplification in our OPA generated via a split-step Fourier method. As pump energy increases, the pump pulse depletes more efficiently given the peak-power-enhanced nonlinear interaction, and causes the peak of $P(N)$ to shift away from 0. Fig. \ref{fig:pump}B shows simulated $P(N)$ with 20-nm bandpass filtering at degeneracy before measurement. In this case, the dominant contribution to the peak shift is pump depletion as multimode effects are suppressed. The difference can be seen in Fig. \ref{fig:pump}A where in the unfiltered case, peak shifting is more prominent thanks to strong multimode contributions. Peak-shifting as a result of pump depletion has recently been studied in single-mode quantum simulations \cite{chinni2024beyond} and experimentally observed \cite{florez2020pump}.

\section{Appendix 4: Overcoming the Electronic Bandwidth Limitation}

While the OPAs demonstrated in this paper can achieve clock speeds of 6.5 THz, our current experimental setup is still constrained to the speed of the electronics used for photodetection. This problem of needing to return to the electronic domain will continue to exist until we use an all-optical computer to process experimental data \cite{li2025all}. To overcome this, we can employ OPAs as fast switches to demultiplex a clock rate of $f$ down to $f/N$ using $N$ OPAs and photodetectors. Figure \ref{fig:multiplex} shows an example demultiplexer circuit design. An unknown quantum state is amplified and measured inside of the measurement OPA. After measurement, we are left with classical information robust to losses which we split into $N$ channels. Once $N$ clock cycles elapse, we have $N$ pulses in each of the $N$ channels. For each channel, a second pump pulse dubbed the "gating pump" is sent in to pump the gating amplifier. To achieve demultiplexing, each gating pump is time delayed such that in channel i, the ith pulse will be amplified in the gating amplifier while all other pulses will remain unamplified. While each of the $N$ photodetectors will see a THz stream of pulses, only one out of every $N$ of these pulses will be amplified enough to produce an electronic signal at the photodetector output. Hence, by correctly setting the delays of the gating pump pulses, we can selectively send each pulse to a different photodetectors and effectively lower the system clock rate by a factor of $N$. The feasibility dispersion-engineered OPAs as switches for ultrafast demultiplexing was first studied in \cite{guo2022femtojoule}, however similar schemes have also been presented elsewhere in literature \cite{li2025all}.

\providecommand{\noopsort}[1]{}\providecommand{\singleletter}[1]{#1}%

\end{document}